# EcoSphere: A Decision-Support Tool for Automated Carbon Emission and Cost Optimization in Sustainable Urban Development

Siavash Ghorbany, sghorban@nd.edu, [0000-0002-9588-0527]
*University of Notre Dame, United States*

Ming Hu, mhu1@nd.edu, [0000-0003-2583-1161]
*University of Notre Dame, United States*

Siyuan Yao, syao2@nd.edu, [0000-0002-4093-193X]
*University of Notre Dame, United States*

Matthew Sisk, msisk1@nd.edu, [0000-0002-4141-9655]
*University of Notre Dame, United States*

Chaoli Wang, chaoli.wang@nd.edu, [0000-0002-0859-3619]
*University of Notre Dame, United States*

**Abstract**
The construction industry is a major contributor to global greenhouse gas emissions, with embodied carbon being a key component. This study develops EcoSphere, an innovative software designed to evaluate and balance embodied and operational carbon emissions with construction and environmental costs in urban planning. Using high-resolution data from the National Structure Inventory, combined with computer vision and natural language processing applied to Google Street View and satellite imagery, EcoSphere categorizes buildings by structural and material characteristics with a bottom-up approach, creating a baseline emissions dataset. By simulating policy scenarios and mitigation strategies, EcoSphere provides policymakers and non-experts with actionable insights for sustainable development in cities and provide them with a vision of the environmental and financial results of their decisions. Case studies in Chicago and Indianapolis showcase how EcoSphere aids in assessing policy impacts on carbon emissions and costs, supporting data-driven progress toward carbon neutrality.

## 1    Introduction

The construction industry generates 40% of global greenhouse gas emissions, with embodied carbon accounting for up to 57% in low-energy buildings and 74–100% in NetZero buildings (University of Maryland, 2021; United Nations Environment Programme, 2022). Building-related $CO_2$ emissions have hit historic highs, 5% above 2020 levels (United Nations Environment Programme, 2022), highlighting the urgent need for industry-wide and urban-scale interventions to address embodied carbon.

Carbon emission reductions extend beyond environmental benefits, positively impacting public health, economic growth, and energy security (Farooq *et al.*, 2019; Jacobson *et al.*, 2019; Lin and Raza, 2020; Chu and Zhao, 2021; Hussain *et al.*, 2022). These benefits have spurred various national and international carbon neutrality initiatives, including the Paris Agreement, which seeks net-zero carbon emissions by mid-century (United Nations, 2015; Wang *et al.*, 2021). However, reducing emissions in urban planning, particularly in developed countries, is challenging due to complex urban infrastructure, economic factors, and resource limitations (Wang and Jiang, 2020; Sharif and Tauqir, 2021). Barriers such as limited high-resolution building data and decision-support tools, especially in the U.S., hinder progress despite the country's potential to significantly impact global emission reductions (Larch and Wanner, 2024).

Current policy-making tools often lack building-level insights, underscoring the need for solutions that enable data-driven decisions. With the objective of moving towards urban sustainability, this study aims to develop a standalone simulation tool applicable to all the cities in the United States to assess the potential scenarios for embodied and operational carbon in these cities. This research introduces EcoSphere, an integrated software solution designed to address the need for detailed, building-level data and provide stakeholders with accessible simulations to guide sustainable urban planning. EcoSphere leverages the National Structure Inventory, combined with computer vision and natural language processing applied to Google Street View and satellite imagery, to create a high-resolution dataset categorizing buildings by material and structure. Its scenario-based simulations enable stakeholders to evaluate the impacts of policy decisions on carbon emissions and costs, offering a robust, accessible tool for sustainable urban planning.

## 2   Literature Review

Data-driven decision-support tools have become increasingly vital for managing carbon emissions in cities, where built environments play a major role in overall greenhouse gas output (Aumann, 2007; Faulin *et al.*, 2010; de Paula Ferreira, Armellini and De Santa-Eulalia, 2020). Two predominant research strands underlie this field. The first emphasizes **predictive models**, using statistical or machine learning methods to forecast emission levels under specific conditions (Saad *et al.*, 2020; Chu and Zhao, 2021; Fang, Lu and Li, 2021; Su *et al.*, 2023). Although these approaches offer initial baselines, they typically lack sufficient granularity to inform building-level interventions (Gao *et al.*, 2023; Hu and Ghorbany, 2024). The second strand harnesses **simulation-based techniques**—for example, modeling hypothetical changes in transportation networks (Gao, Hu and Peng, 2014; Wu and Zhao, 2016) or tracking land expansion (Hu *et al.*, 2022; Wang, Zeng and Chen, 2022; Li *et al.*, 2023; Tian and Zhao, 2024; Wu *et al.*, 2024), to explore "what-if" scenarios. Yet, many of these studies omit the embodied carbon of diverse building stocks or focus narrowly on a single city sector.

Where researchers attempt building-level analyses at scale, top-down estimates often prevail (Hu and Ghorbany, 2024), aggregating emissions while overlooking material, structural, and age differences among individual buildings (Li and Deng, 2023). Conversely, smaller-scale studies may use audits or machine learning to derive operational energy and embodied carbon intensities, but they struggle to expand beyond limited samples (Zhang *et al.*, 2021). This reveals a pressing gap: existing frameworks rarely provide **high-resolution, bottom-up modeling** at a citywide scale (Ghorbany and Hu, 2024). Data constraints compound this challenge: open records such as county assessor database are frequently incomplete, and remote-sensing data must be filtered through specialized computer vision processes (Ghorbany and Hu, 2024; Ghorbany, Hu, Sisk, *et al.*, 2024; Ghorbany, Hu, Yao and Wang, 2024; Ghorbany, Hu, Yao, Wang, *et al.*, 2024; Hu *et al.*, 2025; Yao *et al.*, 2025).

Even when such datasets can be constructed, decision-makers often lack **user-friendly simulation tools** that integrate embodied and operational emissions under various policy scenarios (Hu, 2022; Hu and Ghorbany, 2024). Existing methods typically focus on singular aspects—like renovation costs or demolition rates—without offering a unified framework to capture the interplay of material substitutions, building lifespans, and urban expansion. The outcome is limited practical relevance for city planners, who need both financial and environmental metrics to guide policy.

By developing a **scalable, archetype-based** decision-support platform, the present study addresses these key deficits. Rather than relying on top-down averages, EcoSphere leverages the National

Structure Inventory, georeferenced assessor data, and AI-driven classification of Google Street View and satellite imagery to form a robust baseline of building-level attributes. This **bottom-up methodology** highlights the carbon impact of diverse structural types, foundations, and materials. In turn, it supports customizable simulations of demolition, renovation, and new construction across entire cities, combining **embodied** and **operational** emissions to yield actionable, scenario-specific results. Thus, EcoSphere fills the major gap by offering **finer resolution**, **wider scalability**, and **policy-focused outputs**—features largely absent from current urban carbon modeling frameworks.

# 3 Research Methodology

## 2.1 General Framework

The EcoSphere development framework comprises four stages: data creation, building categorization, scenario-based simulation, and tool development. It models urban building stocks to analyze carbon emissions and costs across policy scenarios, integrating data and simulation into a user-friendly tool to support informed, sustainable urban development decisions.

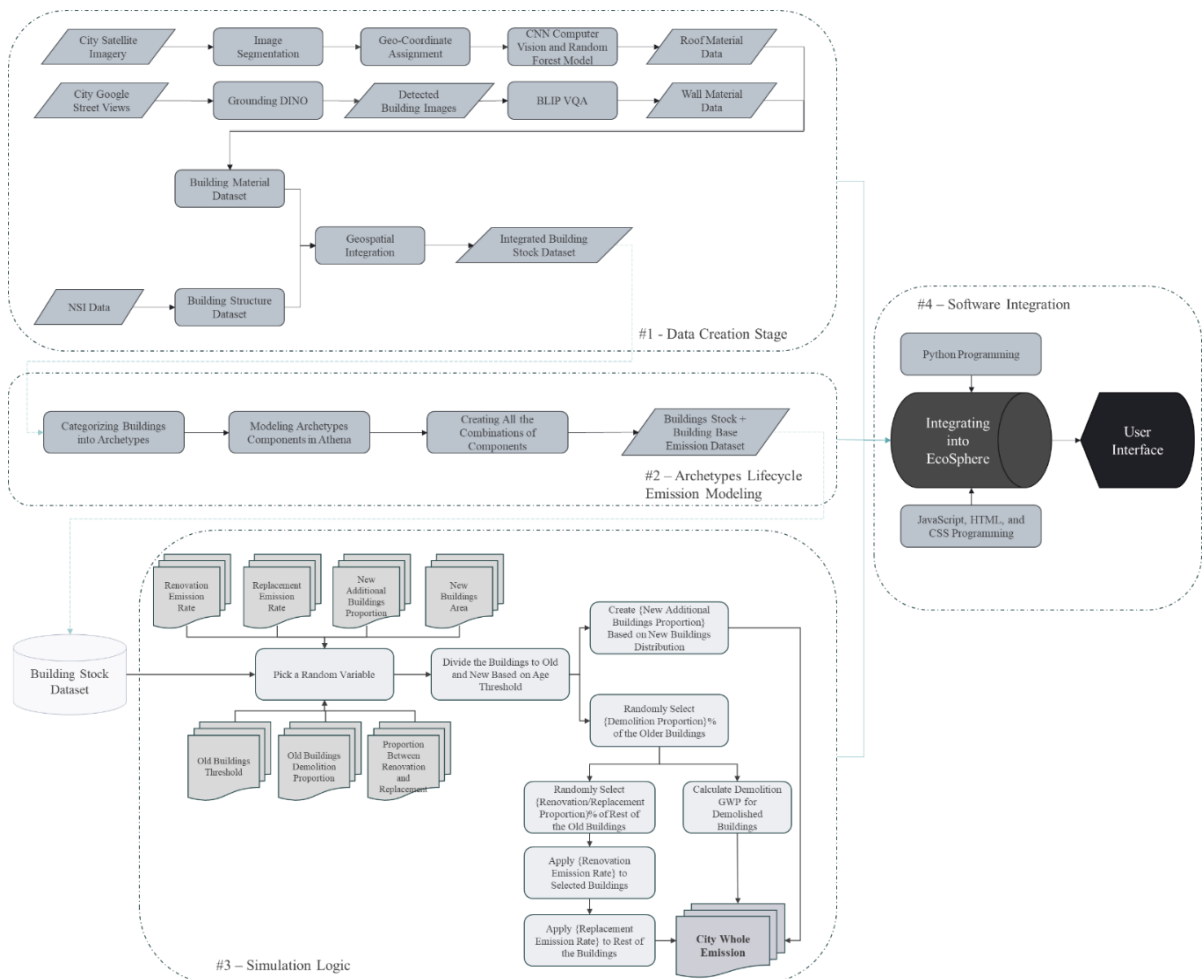

**Figure 1 - The EcoSphere Development Flowchart**

## 2.2 Stage 1: Building Stock Dataset Creation

Developing a robust dataset is essential for emissions modeling. This study employs a bottom-up, archetype-based approach requiring detailed building data, including wall and roof materials, structure, foundation, height, and area (Hu and Ghorbany, 2024). Data collection is challenging due to scattered sources; for instance, Cook County Open Data provides envelope data for Chicago (Cook County

Assessor's Office, 2022), while the NSI includes structure, foundation, and height details (NSI, 2024). Most cities and states lack such granular data.

When data on specific materials is not publicly available, computer vision models were applied to Google Street View (GSV) images and satellite imagery to classify building wall and roof material, following the methodologies and metrics from the authors' previous studies (Ghorbany, Hu, Sisk, *et al.*, 2024; Ghorbany, Hu, Yao, Wang, *et al.*, 2024; Hu *et al.*, 2025; Yao *et al.*, 2025). EcoSphere uses geographical coordinates (latitude and longitude) to integrate data from various sources at the building level, creating a high-resolution geospatial dataset. Pilot studies include over a million buildings in Chicago, 300,000 in Indianapolis, 100,000 in South Bend, and 500,000 in Houston. This dataset establishes a reliable baseline for urban emissions, enhancing EcoSphere's analytical capabilities.

## 2.3 Stage 2: Archetype Emission Dataset Creation

The next stage involves creating an emissions dataset for different building archetypes, which is essential for scalable emissions modeling. Given the sheer volume of buildings, calculating emissions for each structure individually is impractical. Instead, buildings are categorized into archetypes based on their wall material, roof material, structural type, and foundation, with a bottom-up approach (Ghorbany and Hu, 2024; Hu and Ghorbany, 2024). In Chicago, for instance, the 1,010,840 buildings are organized into 157 archetypes, each representing unique combinations of these attributes.

Each archetype is modeled as a standardized building unit (e.g., 1,000 square feet, single-story) using the Athena Impact Estimator for Buildings software, producing emission tables for life cycle stages, particularly embodied carbon. Archetypes are coded as A-B-CC-DD, where A denotes structure type, B foundation type, CC wall material, and DD roof type. For instance, "WBW2R1" represents a wooden structure, basement foundation, masonry wall, and shingle roof. This process generated a dataset of 1,715 unique archetypes for the U.S. By scaling emissions based on building size and floors, EcoSphere estimates emissions across the entire building stock using this archetype baseline. This study considers both residential and commercial buildings, integrated from NSI data, and calculates the operational carbon accordingly and based on buildings' activity type (e.g., educational, hospital, etc.).

## 2.4 Stage 3: Simulation Logic Design

EcoSphere's simulation logic models various urban planning scenarios and their impact on carbon emissions. Python programming, based on a Monte Carlo logic (Ramezani Etedali *et al.*, 2023; Ghorbany and Hu, 2024), was utilized to simulate decision scenarios for buildings based on their age and condition, categorized as new (less than 20 years), mid-range (20-50 years), and old (over 50 years). The number 50 - 80 is the optimal desired lifespan here where the software allows the user to modify it. A city expansion component introduces a 1-5% increase in new construction, further reflecting urban growth dynamics. Table 1 demonstrates the parameters' ranges in the simulation logic.

**Table 1 - Simulation Scenario Parameters**

| Category | Variable | Range |
|---|---|---|
| **Building Age Thresholds** | Building Lifespan Threshold | 50 – 80 years (10-year increment) |
| | Newer Buildings Age Threshold | 20 years |
| **Buildings' Change Parameters** | Older Buildings Demolished Proportion | 20 – 50% (10% increment) |
| | Renovation Emission Rate | 100 – 150% (5% increment) |
| | Replacement Emission Rate | 100 – 300% (10% increment) |
| | Proportion Between Renovation and Replacement | 0.1 – 0.95 (0.05 increment) |
| **New Construction Parameters** | New Additional Buildings Proportion | 1 – 5% (Uniform) |
| | New Buildings Area | 80 – 120% (Uniform) |
| **Cost Values** | Commercial Renovation | $450/ft² |
| | Commercial New Construction | $562/ft² |
| | Commercial Demolition | $10/ft² |
| | Residential Apartment New Construction | $508/ft² |

| | |
|---|---|
| Residential Single-family New Construction | $200/ft² |
| Residential Apartment Renovation | $400/ft² |
| Residential Single-family Renovation | $100/ft² |
| Residential Demolition | $15/ft² |

As part of assumptions, each old building is randomly assigned one of three scenarios: demolition, renovation, or replacement, to reflect possible aging structure paths. Emissions are calculated using archetype-specific baselines, and costs are assigned based on typical expenses, such as $100 per square foot for residential renovations and $562 per square foot for commercial replacements. These default values, shown in Table 3, can be adjusted in EcoSphere's interface for contextual simulations. The simulation also includes six mitigation strategies, such as lifespan extension, space optimization, wood material substitution, recycling enhancement, and prefabrication (Hu, 2022), enabling users to explore carbon footprint reductions. This simulation logic is detailed in stage #3 of Figure 1. Even though these assumptions can impact the outcome of the model, these values are all definable by the user.

## 2.5    Stage 4: Software Development

EcoSphere is developed using Python, JavaScript, HTML, and CSS to deliver a user-friendly interface for urban emissions simulations. Users can adjust parameters, select cities, emission types, building samples, and analysis years, then visualize and save results categorized by factors like building lifespan and policies. Tailored for researchers and policymakers, it enables analysis of entire building stocks or samples for efficiency. The dashboard presents detailed outputs, including cost and emissions analyses across scenarios. Figure 2 showcases EcoSphere's intuitive interface for customizing simulations and conducting tailored analyses.

# 4    Findings and Discussion

EcoSphere's interface (Figure 2) is designed to support both technical users and policymakers by offering an accessible platform to run city-specific simulations of carbon emissions and costs. Users can adjust model parameters, select cities, define building lifespan thresholds, and set emission targets. The simulation then generates detailed results stored in a designated directory, which can be visualized in the dashboard to assist in interpreting outcomes across various scenarios.

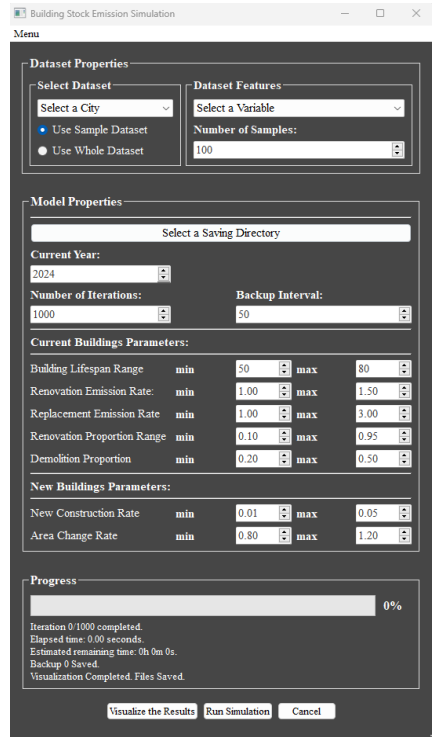

**Figure 2 - The Simulation Tool Interface** (Ghorbany *et al.*, 2025)

The interface's modular design includes sections for adjusting parameters, tracking simulation progress, and displaying results. The **data properties** section allows users to choose the city for analysis and specify the scope (e.g., full city or a randomized sample). Users can also select emissions variables, such as Global Warming Potential across different life cycle phases (e.g., Phases A to C), adding flexibility to explore emissions based on the desired scope and resolution. After setting the parameters, the user initiates the simulation, with progress shown in real-time and results stored as .csv files.

Upon completion of the simulations, the **visualization dashboard** (Figure 3) provides various perspectives, such as cost analysis, scenario-driving variables, and emissions comparisons. The default tab, "Regional Review," offers high-level insights into the simulated outcomes, organized by categories like cost, mitigation strategies, and emissions breakdown by building life cycle phases.

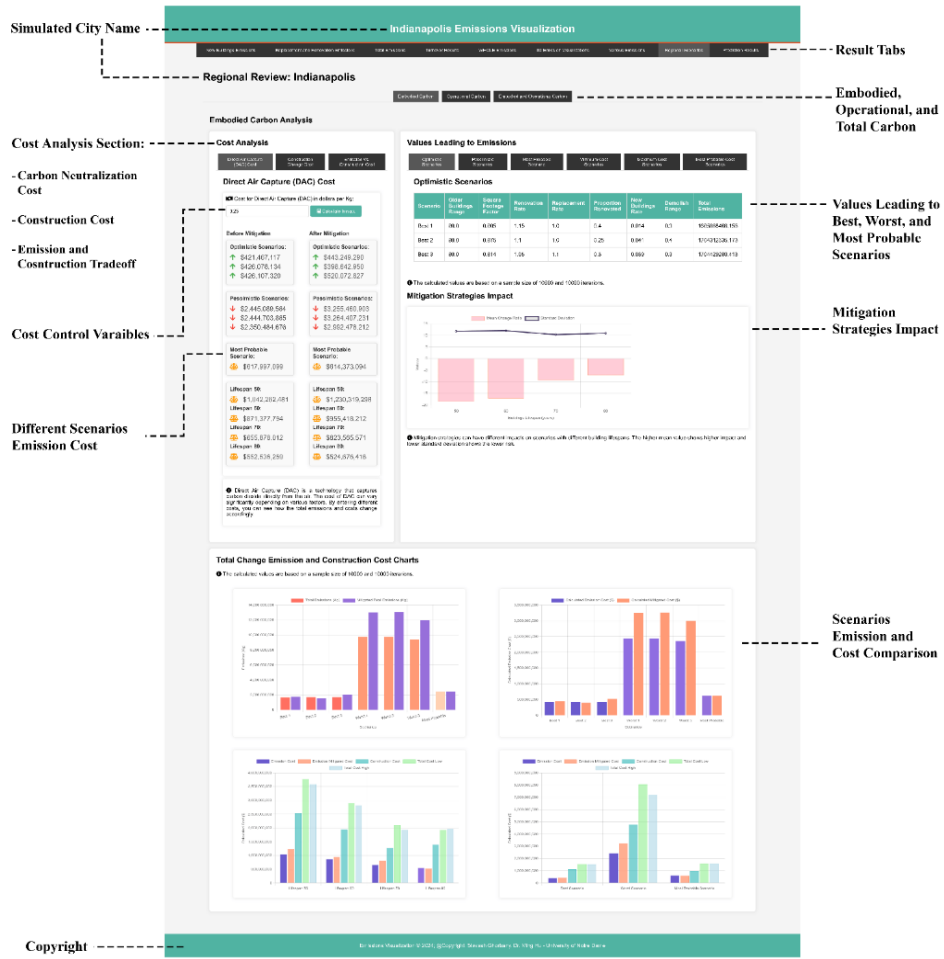

**Figure 3 - The simulation dashboard outcome (the results show a sample of 10600 buildings in the Indianapolis for 10,000 iterations)** (Ghorbany *et al.*, 2025)

EcoSphere's include several sections. The cost analysis section translates emissions into financial terms, aligning carbon-neutrality goals with economic considerations. This section uses the direct air capture (DAC) unit price as the metric for calculating the economic impact of emitted carbon in each scenario. Users can adjust the cost of direct air capture (DAC) to visualize its impact on emissions costs by the selected city. Emissions and their associated DAC costs are calculated across three scenarios: the most probable, the most optimistic, and the most pessimistic outcomes. Users can also view scenario results based on varying building lifespans, allowing them to understand how different initial construction quality phases influence overall costs.

In addition to DAC costs, the cost analysis section includes **Construction Change Cost** and **Emission vs. Construction Cost** sub-sections (see Figure 4). The Construction Change Cost tab allows users to customize unit costs by building and scenario type, while the Emission vs. Construction Cost tab consolidates results from other sections. This summary enables users to quickly assess financial and environmental impacts across different urban policy choices, highlighting potential trade-offs and informing cost-effective decision-making. The construction cost unit prices can be modified by the user in the interface and the default values are aligned with what presented in the Table 1.

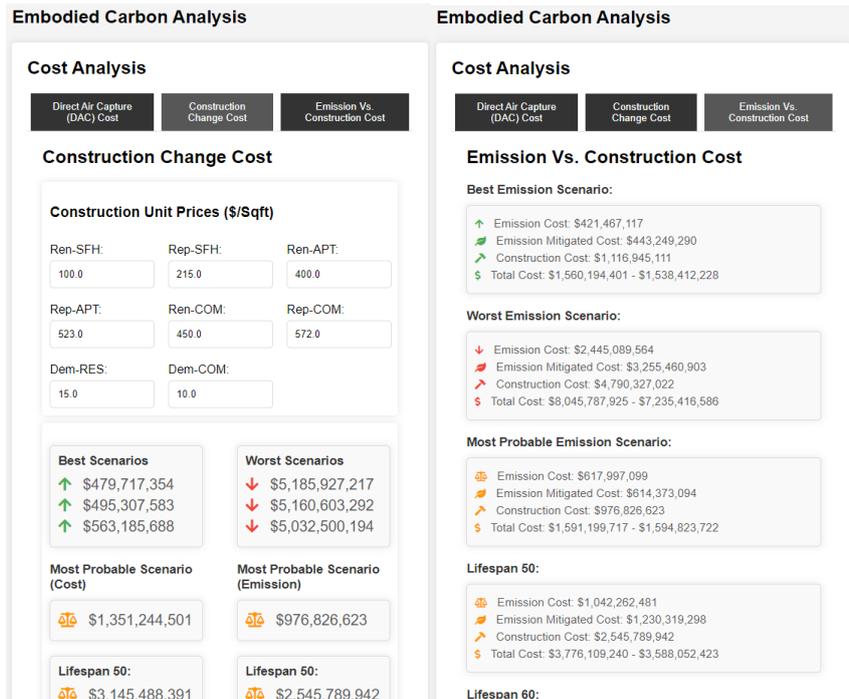

**Figure 4 - Sample of construction change cost and emission vs. construction cost subsections in cost analysis**(Ghorbany *et al.*, 2025)

To offer a deeper understanding of what drives the outcomes, the **Variables Leading to Emissions** section (see Figure 3) highlights factors that contribute significantly to emission levels, such as building lifespan, the demolition, replacement, and renovation rates, and the deployment of six main mitigation strategies. This information provides a detailed explanation of what variables are causing the mentioned emission and costs in the cost analysis section. This tab also displays the number of samples and iterations used, providing context for broader scalability. Users can review the effectiveness of selected mitigation strategies, such as lifespan extension or material substitution, which helps in understanding how these factors influence emissions and costs. This is particularly helpful since some cities can be hugely affected by these strategies while the other cities might already be in their optimal status and further deployment of these strategies can downgrade their status.

EcoSphere's Regional Review section includes visualizations that compare emissions and construction costs across different scenarios, supporting a clear understanding of the data. Figure 5, for example, presents changes in emissions and construction costs across various scenarios for Indianapolis, providing a visual tool to assess the potential impact of specific policies.

Beyond the Regional Review, EcoSphere includes tabs specifically designed for researchers, providing more granular details on emissions. Tabs such as **New Buildings Emissions**, **Replacement and Renovation Emissions**, **Total Emissions**, and **Turnover Results** enable users to view emissions breakdowns by category and probability. For instance, the Total Emissions tab displays a comprehensive overview of emissions for all the buildings in the city and highest probable scenario, while the Turnover Results tab shows scenario-based distribution ratios, illustrating potential shifts in a city's carbon footprint over time (see Figure 6).

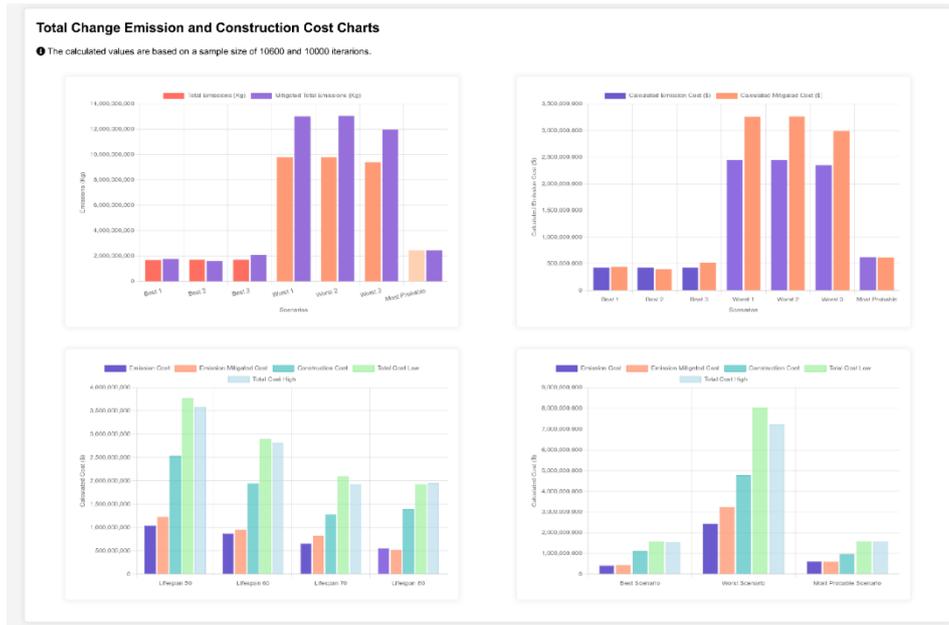

**Figure 5 - Total change emission and construction cost chart section to visualize different scenarios comparison** (Ghorbany *et al.*, 2025)

The Turnover Results section is particularly insightful for policymakers, as it highlights the range of potential outcomes based on city-specific decision paths. For example, results from Indianapolis indicate a range of emissions impacts from 0.9 to 2.2 times current levels. This variation demonstrates the potential of targeted mitigation strategies to reduce emissions, emphasizing the importance of informed planning. This gives a clear understanding of how much investment the city needs. This can also be used later as a benchmark to assess where on the improvement spectrum the city is.

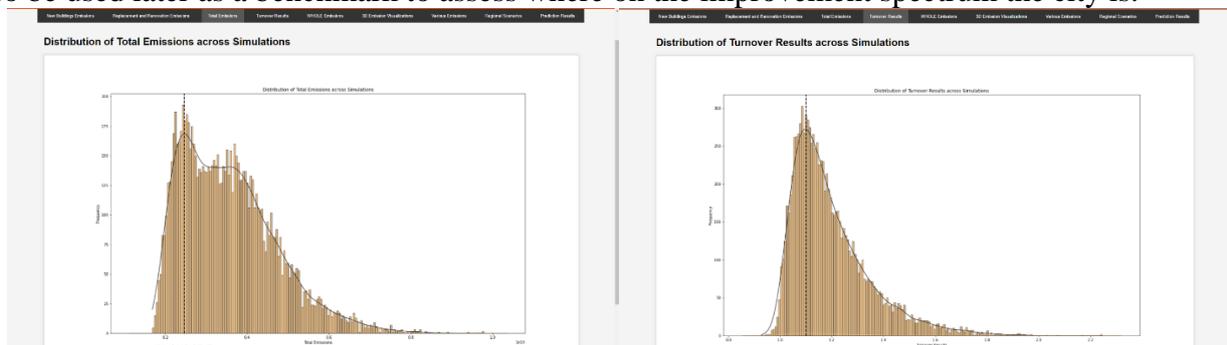

**Figure 6 - Total Emissions Tab to 10600 buildings and 10,000 iterations in Indianapolis (Left), and Turnover results tab for 10600 buildings and 10,000 iterations in Indianapolis (Right)** (Ghorbany *et al.*, 2025)

EcoSphere includes additional tabs for sensitivity analysis, offering insights into both embodied and operational carbon emissions. The **Various Emissions** tab displays emissions for different building lifespans as selected by the user, while the **3D Emissions Visualizations** tab presents a dynamic view of emissions turnover based on demolition rates, building lifespan, and area. These tabs allow users to explore emissions sensitivity to multiple variables, helping to identify key leverage points in policy scenarios.

The final tab in EcoSphere's dashboard features a regression-based prediction model that uses simulation data to project emissions outcomes for different cities. Developed using Ordinary Least Squares (OLS) regression, the model achieves R-squared accuracies of 85-90% depending on the city, demonstrating reliable predictive capability. Users can input specific variables to estimate city emissions, such as changes in demolition rates or building areas, providing a valuable tool for forward-looking urban planning. For example, in Chicago, factors like demolition rates and renovation proportions significantly affect projected emissions, offering a clear roadmap for targeted interventions (refer to Figure 7).

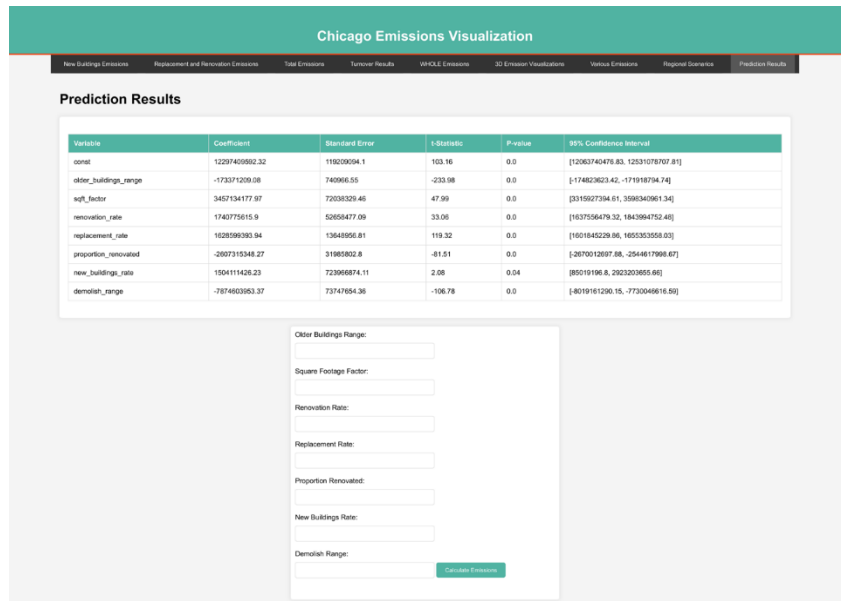

**Figure 7 - The prediction results tab for city of Chicago for 10000 buildings and 12000 iterations** (Ghorbany *et al.*, 2025)

EcoSphere was then piloted in multiple cities, including Chicago and Indianapolis, which share similar Midwest climates, allowing for meaningful comparisons. The results reveal that Indianapolis experiences a narrower range of emissions changes (0.9 to 2.2 times current levels) compared to Chicago (1.8 to 16 times), suggesting that Chicago's higher reliance on masonry structures and high-rise buildings contributes to a larger carbon footprint. Additionally, Indianapolis has implemented more effective mitigation strategies, reflected in its stable emissions patterns.

Indianapolis predominantly uses wood in construction, while Chicago relies on masonry, leading to higher emissions and costs. Optimistic scenarios in both cities show emissions costs near $400 million, while worst-case scenarios reach $2 billion for Indianapolis and $4.5 billion for Chicago. Mitigation strategies could reduce Chicago's emissions costs to $2.9 billion, highlighting the importance of targeted decision-making. EcoSphere's simulations demonstrate the critical role of construction practices and materials in emissions outcomes, highlighting the need for city-specific approaches to carbon neutrality. Indianapolis serves as a model for best practices, while Chicago's results show the potential for improvement through informed policy choices. This emphasizes the importance of the existence of this developed software.

## 5    Conclusions and Further Research

This study addresses the need for effective tools in sustainable urban planning with EcoSphere, an integrated software designed to evaluate embodied and operational carbon emissions in urban environments. EcoSphere combines AI-driven data collection and bottom-up modeling to create a high-resolution dataset of U.S. building stocks, filling a crucial gap in data availability and accessibility for urban-scale emission analysis.

EcoSphere offers an automated solution that enables users to define and assess various urban planning scenarios through a user-friendly interface. By integrating National Structure Inventory data with Google Street View and satellite imagery, the tool builds a comprehensive picture of the urban environment, providing detailed insight into emissions baselines, financial impacts, and potential mitigation strategies. The case studies in Chicago and Indianapolis highlight EcoSphere's ability to capture local variations in construction practices and policies, allowing for tailored emissions reductions.

By focusing on the trade-offs between emissions and costs, EcoSphere serves as a critical decision-support tool. Its intuitive design bridges the gap between complex environmental data and practical urban planning, making sustainability planning accessible for non-experts. Through its visualization dashboard, users can explore potential outcomes and make data-driven decisions that contribute to long-term carbon neutrality goals.

In conclusion, EcoSphere enhances the capability to model and simulate urban carbon emissions, supporting a proactive approach to climate change mitigation. Future work may expand the software's scope to include more detailed operational carbon analyses, integrate renewable energy considerations, and apply the tool across diverse global cities to further validate its effectiveness and flexibility. By advancing automation in urban planning, EcoSphere represents a significant step towards automation in sustainable city development and climate resilience.

Future studies can use this powerful tool and produced dataset through this software simulation to harness the power of advanced statistical methods such as Bayesian Networks to find the causal networks of variables impacting cities' emissions (Ghorbany, Noorzai and Yousefi, 2023; Ghorbany, Yousefi and Noorzai, 2024). Furthermore, statistical analysis of the cities differences can be another research path empowered by this tool. As the main contribution, the EcoSphere removes the carbon analysis scalability challenges and limitations for the United States, allowing for further analysis of different cities and different regions in the U.S. by harnessing the power of advanced AI, Machine Learning, and Computer Vision tools. Future studies could focus on extending EcoSphere to include predictive modeling of urban growth patterns and their impact on emissions and costs. By integrating demographic trends, economic development data, and infrastructure expansion, researchers could simulate how urbanization trajectories influence emissions and explore optimal strategies for sustainable growth.

# 6    Acknowledgement

This study is funded by National Science Foundation #2317971, National Science Foundation #2430623, and University of Notre Dame Lucy Family Institute of Data and Society. We thank all staff in the institute for their contribution to the data collection, sharing, and other support. We thank all faculty, staff, and students in the School of Architecture and Department of Civil Engineering and Computer Science at the University of Notre Dame for their support of this project.